\documentclass[preprint,prb,superscriptaddress,preprintnumbers,amsmath,amssymb]{revtex4}

\usepackage{graphicx}
\usepackage{dcolumn}
\usepackage{bm}

\begin{document}

\title{Isotope effect on electron paramagnetic resonance of boron acceptors in silicon}

\author{A.~R.~Stegner}
\email{stegner@wsi.tum.de} \affiliation{Walter Schottky Institut, Technische Universit\"{a}t
M\"{u}nchen, Am Coulombwall 3, 85748 Garching, Germany}

\author{H. Tezuka}
\affiliation{School of Fundamental Science and Technology, Keio University, Yokohama 223-8522, Japan}

\author{T. Andlauer}
\affiliation{Walter Schottky Institut, Technische Universit\"{a}t
M\"{u}nchen, Am Coulombwall 3, 85748 Garching, Germany}

\author{M. Stutzmann}
\affiliation{Walter Schottky Institut, Technische Universit\"{a}t
M\"{u}nchen, Am Coulombwall 3, 85748 Garching, Germany}

\author{M. L. W. Thewalt}
\affiliation{Department of Physics, Simon Fraser University, Burnaby, British Columbia, Canada V5A 1S6}

\author{M. S. Brandt}
\affiliation{Walter Schottky Institut, Technische Universit\"{a}t
M\"{u}nchen, Am Coulombwall 3, 85748 Garching, Germany}

\author{K. M. Itoh}
\affiliation{School of Fundamental Science and Technology, Keio University, Yokohama 223-8522, Japan}

\date{\today}

\begin{abstract}
The fourfold degeneracy of the boron acceptor ground state in
silicon, which is easily lifted by any symmetry breaking
perturbation, allows for a strong inhomogeneous broadening of the
boron-related electron paramagnetic resonance (EPR) lines, e.g. by
a random distribution of local strains. However, since EPR of
boron acceptors in externally unstrained silicon was reported for
the first time, neither the line shape nor the magnitude of the
residual broadening observed in samples with high crystalline
purity were compatible with the low concentrations of carbon and
oxygen point defects, being the predominant source of random local
strain. Adapting a theoretical model which has been
applied to understand the acceptor ground state splitting in the
absence of a magnetic field as an effect due to the presence of
different silicon isotopes, we show that local fluctuations of the
valence band edge due to different isotopic configurations in the
vicinity of the boron acceptors can quantitatively account for all
inhomogeneous broadening effects in high purity Si with a natural
isotope composition. Our calculations show that such an isotopic perturbation also leads to a shift in the g-value of different boron-related resonances, which we could verify in our experiments. Further, our results provide an independent test and verification of the valence band offsets between the
different Si isotopes determined in previous works.
\end{abstract}

\maketitle

\section{Introduction}
\label{Introduction}
Electron paramagnetic resonance (EPR) and related techniques like
electron nuclear double resonance have contributed extensively to
the understanding of substitutional shallow donors in the different allotropes
of silicon throughout the last 50 years.~\cite{Feher-PRB-I-1959,Feher-PRB-II-1959,Wilson-PRB-1961,Stutzmann-PRB-1987, Mueller-PRB-1999, Huebl-PRL-2006, Pereira-PRB-2009} On the contrary, EPR had
been ineffective for the study of shallow acceptors for a
long time.
The reason behind this asymmetry lies in the different structure of
the respective dopant ground states. While the electronic ground state of substitutional shallow donors in silicon is $s$-like and only twofold spin degenerate, shallow acceptors states have $p$-character and show a fourfold degeneracy.~\cite{Kohn-SSP-1957} This latter degeneracy can
partially be lifted by any symmetry breaking perturbation. Such
perturbations, e.g.~electric fields or strain, can strongly alter
or even dominate the level scheme of acceptor Zeeman energies for external
magnetic fields that are typically used for EPR measurements. If
the perturbation is not homogeneous across the sample, this can
easily lead to an extreme inhomogeneous broadening of the EPR
resonances. Therefore, the observation of a boron-related EPR signal had initially only been possible under application of a strong and homogeneous external stress.~\cite{Feher-PRL-I-1960}

It was only in 1978, when Si became
available with sufficient crystalline quality, that EPR of B
acceptors in externally unstrained Si (Si:B) was reported for the
first time by Neubrand.~\cite{Neubrand-pss86-269-1978,
Neubrand-pss90-301-1978} In particular, a correlation between the
linewidths of the different boron-related EPR resonances and the
concentrations of C and O point defects was established. However, a
number of fundamental questions have remained unsolved: (i)
Although a random strain distribution that is induced by point
defects should lead to a purely Lorentzian broadening of the EPR
resonances, the experimentally obtained lines could only be fitted
with Voigt profiles, taking into account a significant Gaussian
contribution that was found to be independent of the concentrations
of C and O. The origin of this additional broadening, which
dominates the overall linewidths in samples with small
point defect concentrations and which shows a large angular dependence,
has essentially remained unclear. (ii) For the samples of highest
crystalline purity, the Lorentzian contribution to the Voigt
profiles did not fall below a threshold value of 10~mT which, in
the model developed by Neubrand, would correspond to an unreasonably high
concentration of point defects. In
Ref.~\onlinecite{Neubrand-pss90-301-1978}, Si interstitials were
suggested as a possible explanation for this finding. (iii) For two
of the six B-related resonances, a distinct substructure was
observed. It was proposed that this structure originates from a dynamic
effect, however, the specific mechanism remained unknown.~\cite{Neubrand-pss86-269-1978,Lassmann-PRL-1992}

Karaiskaij and coworkers proposed in Ref.~\onlinecite{Karaiskaj-PRL89-016401-2002} that the random distribution of the different stable Si isotopes $^\mathrm{28}$Si, $^\mathrm{29}$Si, and $^\mathrm{30}$Si in Si crystals with a natural isotope composition
plays an important role for the inhomogeneous broadening of
B-related EPR resonances.
In that work, it was shown that the valence band offsets between isotopically pure Si crystals consisting of different isotopes, which lead to local fluctuations of the valence band edge in the vicinity of the different B acceptors in $^\mathrm{nat}$Si, are responsible for the residual ground state splitting of shallow acceptors in this material, which is e.g.~observed in photoluminescence spectra of acceptor-bound excitons.~\cite{Karaiskaj-PRL89-016401-2002,
Karaiskaj-PRL90-016404-2003}
However, it has remained unclear
which of the open questions listed above can be attributed to local valence band fluctuations and whether a quantitative understanding of the
observed effects is possible. In the present paper, we extend the
theoretical model established in
Ref.~\onlinecite{Karaiskaj-PRL90-016404-2003} to explain the
acceptor ground state splitting in the absence of an external
magnetic field ($B$=0) to non-zero magnetic fields and
investigate the influence of isotope-induced perturbations on the line shape
of B-related EPR resonances in Si with the natural and with isotopically engineered isotope compositions. We show that the inhomogeneous broadening effects described in (i) and (ii) can quantitatively be explained by
isotope-induced random local fluctuations of the valence band edge
in the vicinity of the B acceptors. We verify this
finding experimentally via a direct comparison of EPR spectra
obtained from B-doped $^\mathrm{nat}$Si and isotopically
purified B-doped $^\mathrm{28}$Si. The isotope-induced broadening is discussed for different orientations of the magnetic field and different isotope compositions. Additionally, the asymmetric line shape of the B-related EPR resonances observed in $^\mathrm{nat}$Si, which was not addressed in previous work can quantitatively be explained using our model.
We show that the random distribution of the different Si isotopes in $^\mathrm{nat}$Si also shifts the effective g-values of different boron-related EPR resonances with respect to pure $^\mathrm{28}$Si. This effect is quantitatively verified by our experimental data. The crucial parameters for the agreement between this model and the experimental data are the
valence band offsets between pure $^\mathrm{28}$Si,
$^\mathrm{29}$Si, and $^\mathrm{30}$Si, which were extracted from
calculations of the temperature dependence of electronic band
states.~\cite{Cardona-PEP-1989,Karaiskaj-PRL90-016404-2003} Using
our model to recalculate the residual acceptor ground state
splitting for $B$=0, we show that the 30~\% discrepancy between
theory and the experimental value determined by phonon absorption
spectroscopy in
Ref.~\onlinecite{Karaiskaj-PRL90-016404-2003} is not due to an
uncertainty in the values assumed for the valence band offsets, but
rather due to an inaccuracy of the acceptor wave function used in that publication. In our calculations, the best agreement with the experimental data is obtained when the assumed valence band offset between $^\mathrm{28}$Si and $^\mathrm{29}$Si is reduced by 8~\% from the value determined in Refs.~\onlinecite{Karaiskaj-PRL90-016404-2003, Cardona-PEP-1989}.
\section{Theoretical model}
\label{SectionII}
\begin{figure}
\includegraphics[width=8cm]{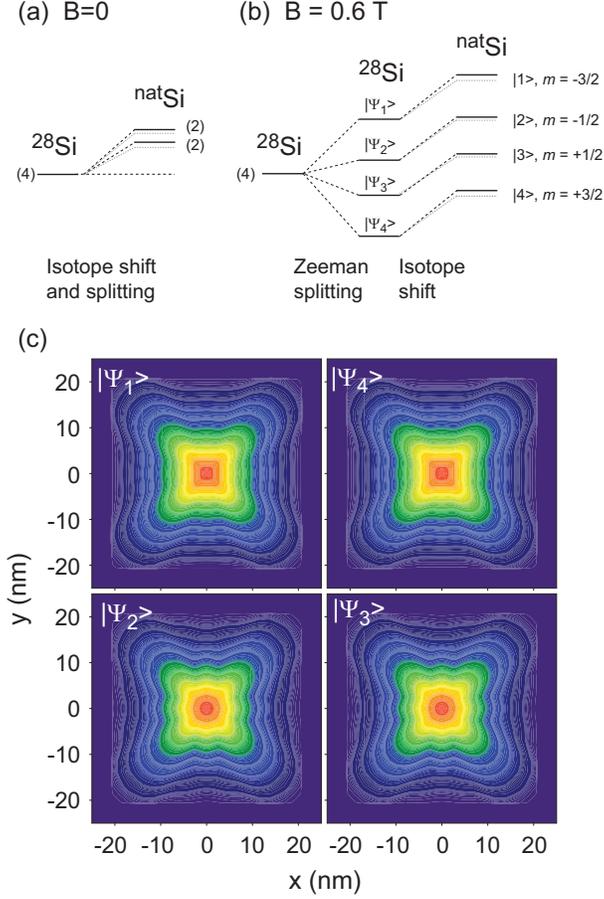}
\caption{\label{figure1} (Color online only) (a,b) Energy level schemes of the acceptor ground state. (a) In the absence of an external magnetic field
$B$, the fourfold degenerate ground state in $^\mathrm{28}$Si is split into two twofold degenerate states in $^\mathrm{nat}$Si both shifted to
higher energies. (b) When a magnetic field of e.g. $B=0.6$~T is
applied, the degeneracy of the ground state is already fully lifted by the Zeeman coupling in $^\mathrm{28}$Si. The isotopic effect leads to an additional
upwards shift of the energy levels which, for a given isotopic
configuration, slightly varies for the different Zeeman levels.
(c) Two-dimensional cross sections at $z$=0 (acceptor situated at the origin) of the envelope
functions of the acceptor ground state probability for the four Zeeman
states shown on a logarithmic scale.}
\end{figure}

Boron acceptors in Si are conventionally described in the
hydrogenic impurity model, where a hole is bound by
the electrostatic potential of the singly charged B-ion which is
screened by the dielectric constant of Si.~\cite{Kohn-SSP-1957} The ground
state envelope function of the hole is the 1$s$ state of a
hydrogen-like atom. As shown in Fig.~\ref{figure1}(a) for B in pure $^\mathrm{28}$Si, this ground state is fourfold degenerate at $B$=0 due to the spin and Bloch basis angular momentum.~\cite{Kohn-SSP-1957} This degeneracy is partially lifted in $^\mathrm{nat}$Si due to the
action of isotope-induced local fluctuations of the valence band
edge in the spatial vicinity of the different B acceptor nuclei which constitute a symmetry breaking perturbation for the acceptor wave function. These fluctuations result from the dependence of the
zero-point vibrational energy on the isotope mass in conjunction with the
renormalization of the electronic states by electron-phonon
interaction.~\cite{Karaiskaj-PRL90-016404-2003} In the ensemble
average, the resulting two Kramers degenerate levels are shifted by
an energy that equals the shift of the valence band edge from
$^\mathrm{28}$Si to $^\mathrm{nat}$Si. However, for different B acceptors, the shift of the different energy levels will undergo slight variations depending on the specific distribution of Si isotopes in the vicinity of the
acceptor nucleus.

When a strong external magnetic field is applied, the situation is different, as shown in Fig.~\ref{figure1}(b). Already the Zeeman interaction between the
acceptor-bound holes and the external magnetic field leads to a
full lifting of the degeneracy of the heavy and
light hole states and the energy eigenstates can approximately be
classified in the total angular momentum basis $|j$=3/2$, m
\in \{ \pm 1/2, \pm 3/2 \} \rangle$, where $j$ and $m$ are the angular momentum and the magnetic quantum number, respectively. For strong enough
external magnetic fields, the isotope-induced effects
can then be treated as a small perturbation. As schematically
shown in Fig.~\ref{figure1}(b), on average this perturbation
should lead to an upwards shift of the four Zeeman levels in energy
similar to the situation shown in Fig.~\ref{figure1}(a) for $B$=0. The assumption that this picture holds for
X-band EPR conditions is motivated by the experimental evidence
that the measured electronic g-values of B in $^\mathrm{nat}$Si
are predominantly determined by the Zeeman interaction in samples
with a low concentration of point
defects.~\cite{Neubrand-pss86-269-1978} The isotopic disorder
therefore only leads to a comparatively small perturbation that mostly
manifests itself in a change of the linewidths and lineshapes of
the different boron-related EPR resonances.

To assess the nature and the magnitude of these isotope-induced
changes, we implement our theoretical model in the following way:
First, we calculate the four Zeeman levels of the acceptor ground
state using a $\mathbf{k \cdot p}$ envelope function
model taking into account all six (2$\times$3) spin-resolved valence bands including the split-off band. The Hamiltonian of the acceptor-bound hole can
schematically be written in the form
\begin{equation}
\label{equation1}
  \hat H = \hat H^{6 \times 6}_\mathbf{k \cdot p} + \frac{g_0 \mu_\mathrm{B}}{2}
  \mathbf{\hat{S}}^{6 \times 6} \mathbf{\cdot B} + V(r),
\end{equation}
which was solved on a real-space grid using the NextNano++
code.~\cite{Birner-IEEE-2007} The first term on the right hand
side represents the six-band effective mass Hamiltonian in a
discrete real-space basis including the magnetic field
$\mathbf{B}$ in a non-perturbative and gauge-invariant
manner, with $\mathbf{B}$ only appearing in phase factors. The
second term, where
$\mu_\mathrm{B}$ is the Bohr magneton, $g_0$ is the free electron
g-value and the $6 \times 6$ spin matrices $\hat S_k$=$1^{3
\times 3}\otimes \hat \sigma_k (k \epsilon \{x,y,z\})$ are
determined by the Pauli matrices $\hat \sigma_k$, couples the spin to the magnetic field. The impurity nucleus is represented by a negative charge at the center of the simulation domain screened by the bulk silicon dielectric
constant of $\varepsilon_\mathrm{Si}$=11.7. The
potential energy of the acceptor hole is given by
\begin{equation}
\label{equation2}
  V(r) = \frac{e^2}{\varepsilon_\mathrm{Si} r} + W(r),
\end{equation}
where we have included the so-called central cell correction
$W(r)$. It phenomenologically corrects for deviations of $V(r)$
from a purely hydrogenic potential close to the impurity nucleus
due to changes in the dielectric properties in this spatial
region.~\cite{Pantelides-PRB-1974} To account for $W(r)$, we adopt the
parametrization
\begin{equation}
\label{equation3}
  W(r) = \frac{e^2}{r}\left[A \mathrm{e}^{(-\alpha r)} + (1-A) \mathrm{e}^{(-\beta r)}- \frac{\mathrm{e}^{(-\gamma
  r)}}{\varepsilon_\mathrm{Si}}\right]
\end{equation}
used in Ref.~\onlinecite{Belyakov-JPHYS-2007}, where the parameters $\alpha$=0.755/$a_\mathrm{B}$,
$\beta$=0.35/$a_\mathrm{B}$, $\gamma$=2.45/$a_\mathrm{B}$, and
$A$=1.14 are phenomenological fitting
parameters and $a_\mathrm{B}$ is the Bohr radius of the effective mass acceptor.~\cite{Belyakov-JPHYS-2007} In order to avoid a
singularity of $V(r)$ at $r$=0, the Coulomb potential is replaced
by $Q \delta_{r,0}$ for the central grid node following
Ref.~\onlinecite{Belyakov-JPHYS-2007}.
The only free parameter in our model, $Q$, has been chosen to reproduce the correct mixing of
the different valence band states in $|\Psi_i\rangle$, which is
responsible for the inhomogeneous energy splitting between the
Zeeman levels in the absence of any further perturbations. A
direct experimental measure for this mixing is the difference
between the transition energies of the transitions $|\Psi_1\rangle$$\leftrightarrow$$|\Psi_2\rangle$ and $|\Psi_2\rangle$$\leftrightarrow$$|\Psi_3\rangle$. This energy difference can directly be calculated from the difference in the resonance fields of the broad and the narrow $\Delta m$=1 EPR lines observed in $^\mathrm{28}$Si. We have fitted $Q$ to reproduce the experimentally observed splitting of 49~mT for $B$$||$[001] (cf. Fig.~\ref{figure3}(b), Sec.~\ref{resultsanddiscussion}). For all further calculations, $Q$ has been fixed to this value.
In order to account for the long range character
of $V(r)$ as well as the strong confinement at the acceptor
nucleus, we use a simulation domain size of
50$\times$50$\times$50~nm$^3$ and an inhomogeneous grid with a
strong concentration of nodes close to the acceptor position, respectively. The wave functions were forced to zero at the boundaries of the simulation domain. With this model, we have calculated the four ground state Zeeman levels
labeled by $|\Psi_1\rangle$, $|\Psi_2\rangle$, $|\Psi_3\rangle$,
and $|\Psi_4\rangle$ (corresponding to $m$=$-3/2$, $m$=$-1/2$, $m$=$+1/2$, and $m$=$+3/2$, respectively), as shown in Fig.~\ref{figure1}(b), for
external magnetic fields with different field strengths and orientations relative to
the crystallographic axes. In Fig.~\ref{figure1}(c), we exemplarily
show typical wave functions obtained by these calculations for $B$=0.6~T with
$B$=$B_z$$||$[001] as two-dimensional cross sections through the
different Zeeman states at $z$=0 and the B nucleus located at the origin. The band warping of the
different valence bands is reflected in the structure of the
envelope functions. While the two heavy hole- and light hole-like
states are very similar to each other, between them there are
distinct differences. From this it already becomes clear that the
zeroth order effect of any perturbation will be an energy shift of
the inner two light hole-like states $|\Psi_2\rangle$ and $|\Psi_3\rangle$ with respect to the outer two heavy hole-like states $|\Psi_1\rangle$ and
$|\Psi_4\rangle$ as schematically indicated in Fig.~\ref{figure1}(b).

Based on the calculated wave functions, the isotope-induced
changes of the band edges are considered perturbatively. For this,
we map the wave functions on a discrete atomistic silicon lattice.
The values of the acceptor envelope functions $\Psi_i(n)$ at the
position of the $n$-th silicon site are evaluated by linear
interpolation. For a silicon lattice that comprises
42$\times$42$\times$42 unit cells, corresponding to the size of the simulation domain, we first choose random configurations of $^\mathrm{28}$Si, $^\mathrm{29}$Si, and $^\mathrm{30}$Si isotopes, which on average corresponds to the
isotope composition intended for the different calculations, e.g. 92.23~\% $^\mathrm{28}$Si, 4.67~\% $^\mathrm{29}$Si, and 3.1~\% $^\mathrm{30}$Si for natural Si. Following
Ref.~\onlinecite{Karaiskaj-PRL90-016404-2003}, we introduce the
isotopic perturbation potential
\begin{equation}
\label{equation4}
  V_\mathrm{iso}(n) = \left\{ \begin{array}{l}
 0\,\,\,\,\,\,\,\,\,\,\,\,\,\,\,\mathrm{for}\,\,\,\,^\mathrm{28}\mathrm{Si}  \\
 \Delta E^\mathrm{29} \,\,\,\mathrm{for}\,\,\,\,^\mathrm{29}\mathrm{Si}  \\
  \Delta E^\mathrm{30} \,\,\,\mathrm{for}\,\,\,\,^\mathrm{30}\mathrm{Si}  \\
 \end{array} \right. .
\end{equation}
The valence band offsets $\Delta E^\mathrm{29}$=0.74~meV and
$\Delta E^\mathrm{30}$=1.46~meV are taken relative to the
$^\mathrm{28}$Si valence band edge and were deduced from
calculations of the temperature dependence of the electronic band
states in Si.~\cite{Karaiskaj-PRL90-016404-2003,Cardona-PEP-1989}
Next, we project the diagonal perturbation potential into the
subspace spanned by the four Zeeman states $|\Psi_i \rangle$,
leading to a 4$\times$4 perturbation Hamiltonian $\hat
H_\mathrm{iso}^{4\times4}$ with the matrix elements
\begin{equation}
\label{equation5}
  H_\mathrm{iso}^{ij} =
  \sum \limits_n
  \langle\Psi_i(n)|V_\mathrm{iso}(n)|\Psi_j(n)\rangle.
\end{equation}
Finally, we diagonalize the total Hamiltonian $\hat H_\mathrm{tot} = \hat
H_\mathbf{J \cdot B }^{4\times4} + \hat H_\mathrm{iso}^{4\times4}$
including the diagonal Zeeman term $\hat H_\mathbf{J \cdot B }^{4\times4} =
E_i \delta_{ij}$, with $E_i$ being the energy of the $i$-th Zeeman
level determined from the $\mathbf{k \cdot p}$ model described above. The isotopic contribution $\hat H_\mathrm{iso}^{4\times4}$
leads to a mixing of the four Zeeman levels and to an upward shift of the states, as depicted in Fig.~\ref{figure1}(b). The relative alignment of the four resulting states $|1\rangle$, $|2\rangle$, $|3\rangle$, and $|4\rangle$ (cf. Fig.~\ref{figure1}(b)) strongly depends on the specific
isotopic configuration in the vicinity of the acceptor nucleus. We
have therefore performed the calculation of the isotope shifts as described above for 200,000 different random isotope configurations, taking into account different magnetic fields, isotope compositions, and isotopic perturbation potentials. The resulting statistical distributions of EPR
transition energies allow the quantitative comparison with the experimental data presented below.

\section{Experimental details}
The $^{\mathrm{nat}}$Si:B sample studied has a B doping concentration of 10$^{14}$~cm$^{-3}$, was grown using the float-zone technique, has a rectangular shape and a size of approximately $3\times 3\times 9$~mm$^{3}$.
The isotopically purified sample, referred to as $^{28}$Si:B in
the text, comes from the neck region of a float-zone crystal and has an enrichment of 99.98~\% $^{28}$Si. It is doped with
boron to a concentration of
3$(\pm 1) \times$10$^{14}$~cm$^{-3}$, has a cylindrical shape with a length of
10~mm in [100] direction and a diameter of $\approx$3~mm.

EPR measurements were performed with a Bruker Elexsys E 500
spectrometer in conjunction with a super-high-$Q$ resonator
(ER-4122SHQE) operated at an X-band microwave frequency of $\nu$$\approx$9.4~GHz.
During the measurement, the samples were cooled to temperatures of typically 3~K using an Oxford ESR 900 helium-flow cryostat.

\section{Results and Discussion}
\label{resultsanddiscussion}
\begin{figure}
\includegraphics[width=7cm]{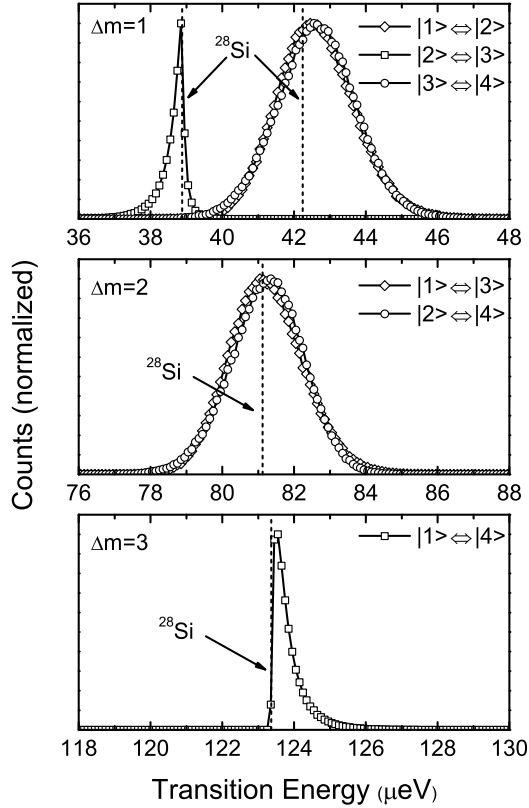}
\caption{\label{figure2}Statistical distribution of the transition
energies between the different Zeeman levels for $^\mathrm{nat}$Si. The transitions are
grouped according to the change in the magnetic quantum number
$\Delta m$. The dashed lines show the transition energies for pure $^{28}$Si, as obtained from our $\mathbf{k \cdot p}$ model without isotopic perturbation. The calculations were performed for an external magnetic field $B$=0.6~T oriented along the [001] axis.}
\end{figure}

In Fig.~\ref{figure2} the statistical distributions that result
from the calculations described above are shown for the six
different EPR transitions in $^{\mathrm{nat}}$Si for $B||[001]$. We find that the
distribution of the inner $\Delta m$=1 transition
($|2\rangle$$\leftrightarrow$$|3\rangle$) and the $\Delta m$=3
transition ($|1\rangle$$\leftrightarrow$$|4\rangle$) are relatively sharp, while the outer $\Delta m$=1 ($|1\rangle$$\leftrightarrow$$|2\rangle$ and $|3\rangle$$\leftrightarrow$$|4\rangle$) and
the $\Delta m$=2 ($|1\rangle$$\leftrightarrow$$|3\rangle$ and $|2\rangle$$\leftrightarrow$$|4\rangle$) transitions show a full width at half maximum (FWHM) of approximately 10~\%
and 5~\% of their mean value, respectively. This can be understood
from the fact that the two former transitions occur between two
states that are dominantly light holes or heavy holes, respectively, and experience a similar shift due to the isotopic perturbation. These transitions are therefore called intrasubband transitions. In
contrast, the latter intersubband transitions comprise both light hole-like and heavy hole-like states that are differently perturbed by the
random isotopic distribution. The asymmetric broadening
of the $|1\rangle$$\leftrightarrow$$|4\rangle$ and the
$|2\rangle$$\leftrightarrow$$|3\rangle$ transitions results from
the mixing of the four unperturbed eigenstates $|\Psi_i\rangle$
that is obtained by the diagonalization of the total
Hamiltonian $\hat H_\mathrm{tot}$. As can be seen, the statistical distribution
of the two outer $\Delta m$=1 transitions and the $\Delta m$=2
transitions are also slightly asymmetric and do not exactly
overlap with each other. For low temperatures, when spin
polarization becomes significant and the equilibrium populations
of the four Zeeman states strongly deviate from each other, this
result should be reflected in an increasing asymmetry of the EPR
lineshape. The dashed lines in Fig.~\ref{figure2} show the transition energies for pure $^{28}$Si, i.e. the transition energies between the unperturbed eigenstates $|\Psi_i\rangle$ obtained from our $\mathbf{k \cdot p}$ calculations. For the maxima of the distributions calculated for $^\mathrm{nat}$Si, we observe a shift in energy with respect to the unperturbed case, which is significant for the outer $\Delta m$=1 and the $\Delta m$=2 transitions. The maxima of the $|1\rangle$$\leftrightarrow$$|2\rangle$ and $|3\rangle$$\leftrightarrow$$|4\rangle$ distributions are shifted to higher energies by 0.315~$\mu$eV and 0.378~$\mu$eV with respect to $^{28}$Si, respectively. For the $|1\rangle$$\leftrightarrow$$|3\rangle$ and $|2\rangle$$\leftrightarrow$$|4\rangle$ $\Delta m$=2 transitions, we find shifts of 0.057$\mu$eV and 0.151$\mu$eV, respectively. These shifts manifest themselves in (temperature dependent) differences between the effective g-values of $^{28}$Si and $^\mathrm{nat}$Si, which will be discussed further below.

\begin{figure}
\includegraphics[width=7cm]{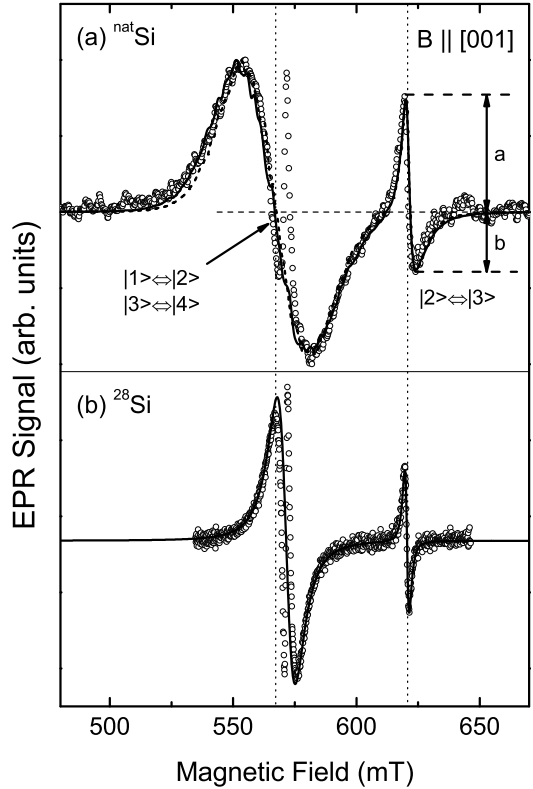}
\caption{\label{figure3}(a) EPR spectrum of the $\Delta m$=1 transitions measured on a $^\mathrm{nat}$Si:B
sample (open circles) shown together with the theoretical
lineshape derived from the distribution of transition energies
shown in Fig.~\ref{figure2} using $\Delta E^\mathrm{29}$=0.74~meV and $\Delta E^\mathrm{30}$=1.46~meV (solid line). The dashed curve shows the simulated spectrum for $\Delta E^\mathrm{29}$=0.68~meV and $\Delta E^\mathrm{30}$=1.34~meV. (b) Corresponding EPR spectrum of isotopically purified $^\mathrm{28}$Si:B. In (b), the solid line is a numerical fit using the first derivative of two Lorentz lines. The orientation of the external magnetic field is $B||[001]$ in both cases. The measurement temperature was 3~K. The dotted lines indicate the resonance fields (zero crossings) of the inner and outer $\Delta m$=1 resonances for $^\mathrm{nat}$Si.}
\end{figure}

\begin{figure}
\includegraphics[width=7cm]{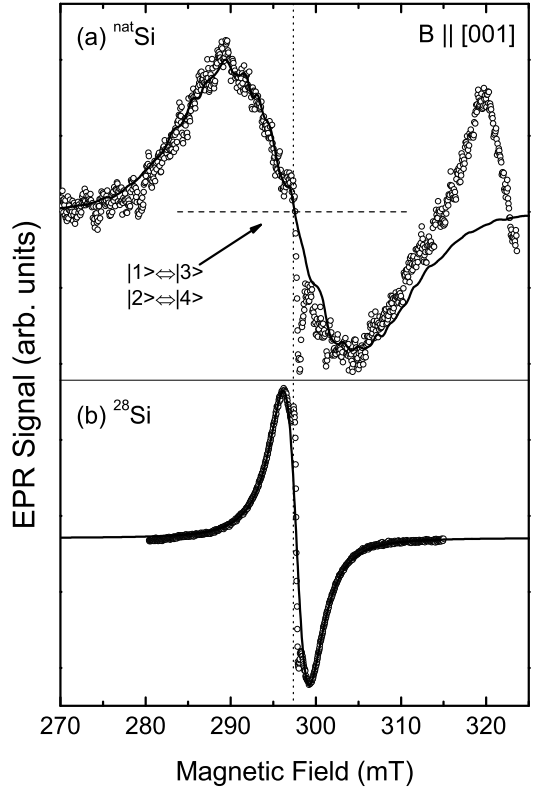}
\caption{\label{figure3b}(a) EPR spectrum of the $\Delta m$=2 transitions measured on a $^\mathrm{nat}$Si:B
sample (open circles) shown together with the theoretical
lineshape derived from the distribution of transition energies
recalculated for $B$=0.3~T using $\Delta E^\mathrm{29}$=0.74~meV and $\Delta E^\mathrm{30}$=1.46~meV. (b) Corresponding EPR spectrum measured on isotopically purified $^\mathrm{28}$Si:B. In (b), the solid line
is a numerical fit using the first derivative of one Lorentz
line. The orientation of the external magnetic field is
$B||[001]$ in both cases. The measurement temperature was 3~K. The dotted line indicates the zero crossing of the $\Delta m$=2 resonance for $^\mathrm{nat}$Si.}
\end{figure}

In order to see to which extent the properties of boron EPR spectra that
have not been understood so far can be explained by the calculated
distributions of the transition energies, we directly compare our
results with experimental data. The open circles in
Fig.~\ref{figure3}(a) show a typical EPR spectrum of the $\Delta
m$=1 transitions measured on a $^\mathrm{nat}$Si sample at
$B||[001]$. The spectrum shows a broad structured line at
$B$=567.2~mT, which is attributed to a superposition of the
resonances originating from the
$|1\rangle$$\leftrightarrow$$|2\rangle$ and the
$|3\rangle$$\leftrightarrow$$|4\rangle$
transitions.~\cite{Neubrand-pss86-269-1978} Further, we observe a
narrower asymmetric resonance at $B$=621.1~mT, which can be
assigned to the $|2\rangle$$\leftrightarrow$$|3\rangle$
transition.~\cite{Neubrand-pss86-269-1978} The origin of the
narrow substructure feature centered at $B$=570.2~mT and discussed under (iii) in Sec.~\ref{Introduction} is the result of an interplay
between spin excitations and spin relaxations in the four level spin
system and is the subject of a different publication.~\cite{Tezuka-2009} In Fig.~\ref{figure3b}(a), the open circles show the EPR spectrum measured on the same $^\mathrm{nat}$Si sample in a magnetic field range where the $\Delta m$=2 resonances are expected. The broad peak observed at $B$=297.5~mT can be assigned to a superposition of the $|1\rangle$$\leftrightarrow$$|3\rangle$ and $|2\rangle$$\leftrightarrow$$|4\rangle$ resonances.~\cite{Neubrand-pss86-269-1978} Here, the narrow substructure line has the opposite sign as compared to the substructure of the broad $\Delta m$=1 peak. The additional signal in the range from $B$=314~mT to $B$=324~mT is not related to boron but is due to a background signal of our $^\mathrm{nat}$Si sample which is of unknown origin.

The curves (b) in Figs.~\ref{figure3} and \ref{figure3b} show the corresponding spectra that were
measured on the $^\mathrm{28}$Si sample. We observe the same
features as in the spectra of the $^\mathrm{nat}$Si sample,
however, the linewidths and the line shapes of all resonances are
strongly changed. The solid lines in Figs.~\ref{figure3}(b) and \ref{figure3b}(b) show computer simulations of the spectra using single Lorentzian lines for each resonance peak,
neglecting the narrow substructures. The best fit which leads to a very good agreement with the experimental data, is obtained for peak-to-peak linewidths of 7.2~mT and 1.7~mT
for the low-field and the high-field $\Delta m$=1 resonances, respectively. For the $\Delta m$=2 resonance, a linewidth of 3.1~mT is obtained. In
spite of the Lorentzian line shapes, we conclude from the temperature
dependence of the linewidths (not shown) that the $|1\rangle$$\leftrightarrow$$|2\rangle$,
$|3\rangle$$\leftrightarrow$$|4\rangle$ resonance and the $\Delta m$=2 resonance shown are inhomogeneously broadened, since the linewidths of these two resonances become almost temperature independent below 3~K.
We attribute this broadening to a distribution of the transition energies between the different energy levels induced by random local strains due to C, O, and B point defects.~\cite{Neubrand-pss90-301-1978,Bir-JPCS-1963} A quantitative estimation of the point defect concentration is given later in the text. For the narrow $\Delta m$=1 resonance at higher magnetic fields, belonging to the $|2\rangle$$\leftrightarrow$$|3\rangle$ transition, the strain induced broadening is only a second order perturbation. For this resonance, the linewidth continues to decrease for temperatures down to 0.3~K and at the temperature investigated here (3~K), the linewidth of this resonance is therefore a good estimate for the natural linewidth.

Therefore, we use the line shape of the $|2\rangle$$\leftrightarrow$$|3\rangle$
resonance in $^\mathrm{28}$Si for a convolution with the
statistical distribution of spin packets shown in
Fig.~\ref{figure2} in order to obtain simulated line shapes for
$^\mathrm{nat}$Si that can be compared to the
experimental spectra of the $\Delta m$=1 and $\Delta m$=2 resonances in
Fig.~\ref{figure3}(a) and Fig.~\ref{figure3b}(a), respectively. To make a direct comparison of the line shapes easier, the calculated spectra have been shifted rigidly on the magnetic field axis so that they overlap with the experimental data. This shift corrects for the inaccuracy of the absolute g-values obtained from our calculations. However, we only performed a rigid shift of the entire spectrum by a constant field $\Delta B$ and the splitting of the $\Delta m$=1 resonances has not been changed. As our model does not yield EPR transition probabilities, the amplitudes of the simulated peaks have manually been adjusted to the measured values. Neglecting the substructure of the broad transitions caused by dynamic effects,~\cite{Tezuka-2009} the calculated signals (black solid curves) show a striking agreement with the experimental spectra, which demonstrates that all inhomogeneous broadening effects
in our $^\mathrm{nat}$Si sample can be explained by isotope-induced fluctuations of the valence band edge
alone. The apparent noise in the simulated spectra is the result of the still somewhat limited number of random isotope distributions used for the simulation. As indicated by the dotted vertical lines, also the shifts in the effective g-values between $^\mathrm{28}$Si and $^\mathrm{nat}$Si that have been predicted as shifts in the transition energies in Fig.~\ref{figure2}, are found experimentally. For the broad $\Delta m$=1 resonance in Fig.~\ref{figure3}, we observe a shift of 4.2~mT which translates into a shift in transition energy of 0.3~$\mu$eV at $B$=0.6~T and is in good agreement with the theoretical values. For the narrow $\Delta m$=1 resonance, we do not see a change in transition energy within the resolution of our measurement, as expected from our calculations. The $\Delta m$=2 resonance shows a smaller shift of 0.5~mT corresponding to a change in transition energy of 0.1~$\mu$eV at $B$=0.6~T.
We note that the resonance fields of the substructures, both for the $\Delta m$=1 and $\Delta m$=2 transitions, are not shifted due to the isotopic perturbation, which means that the substructure lines are not located exactly at the maximum of the broader absorption lines but are slightly shifted to higher fields in $^\mathrm{nat}$Si.

We note that the narrow substructures on the broad $\Delta m$=1 resonance showing a negative sign and the $\Delta m$=2 resonance showing a positive sign for the temperature investigated here, do not originate from a superposition of two absorption peaks which are shifted with respect to each other on the magnetic field axis. Such a splitting of resonance lines could in principle result from slightly different g-values of the two boron isotopes $^\mathrm{10}$B and $^\mathrm{11}$B, which have a slightly different binding energy (1.9$\times$10$^{-2}$~meV higher for $^\mathrm{10}$B).~\cite{Cardona-RMP-2005} However, due to their natural abundance of 19.9~\% and 80.1~\%, respectively, such an effect has to lead to an asymmetric resonance line, which contradicts the highly symmetric line shape of the B-related resonances observed in $^\mathrm{28}$Si, and can therefore be excluded. Another effect that in principle has to be considered for the understanding of the line shape is the hyperfine interaction of the acceptor holes with the nuclear spins of the corresponding acceptor nuclei and $^\mathrm{29}$Si ligands. For $^\mathrm{10}$B and $^\mathrm{11}$B, the nuclear spin is 3 and 3/2, respectively. If the acceptor hyperfine splitting could be resolved in our spectra, we would therefore expect a multiplet of lines rather than a twofold splitting of the resonance line that might be invoked to account for the central substructures. We neither observe a resolved hyperfine signature in any of our spectra nor do we need a further source of inhomogeneous broadening for the description of our spectra that could be attributed to unresolved B hyperfine multiplets or a strong superhyperfine interaction with $^\mathrm{29}$Si ligands. We conclude that the hyperfine coupling strength of B acceptors to their nuclei is much smaller than for the case of shallow donors in Si. This is a result of the $p$-character of valence band Bloch states, which cancels the strong Fermi contact hyperfine term at the position of the dopant atom due to a vanishing probability amplitude of the acceptor hole at its nucleus. For heavy hole states in III-V semiconductor quantum dots, it has recently been pointed out that the anisotropic dipolar hyperfine coupling strength can amount to up to $\approx$10~\% of the Fermi contact term of electrons.~\cite{Fischer-ARXIV-2008} However, it is not necessary to take into account an additional dipolar hyperfine or superhyperfine term for the description of our data.

\begin{figure}
\includegraphics[width=7cm]{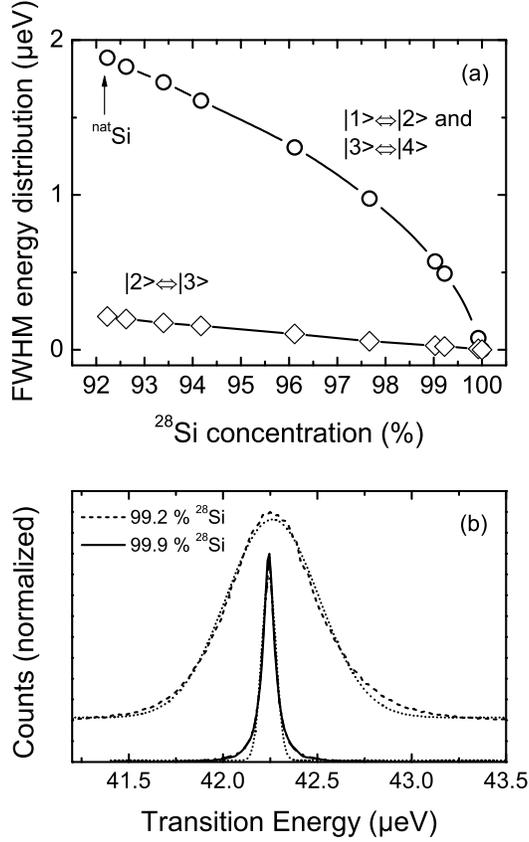}
\caption{\label{figure2b} (a) FWHM of the sum of the transition energy distributions of the $|1\rangle$$\leftrightarrow$$|2\rangle$ and $|3\rangle$$\leftrightarrow$$|4\rangle$ transitions (circles), and the distribution of the $|2\rangle$$\leftrightarrow$$|3\rangle$ transition energies (diamonds) as a function of the $^\mathrm{28}$Si concentration. The solid lines are guides to the eye. The calculations where performed for an external magnetic field $B$ oriented along the [001] axis. The ratio between the residual $^\mathrm{29}$Si and
$^\mathrm{30}$Si concentrations was kept constant at the
value of $^\mathrm{nat}$Si. (b) Distribution of the $|1\rangle$$\leftrightarrow$$|2\rangle$, $|3\rangle$$\leftrightarrow$$|4\rangle$ transition energies for a $^\mathrm{28}$Si concentration of 99.2~\% (dashed line) and 99.9~\% (solid line). The dotted lines are numerical fits using a Gaussian line shape. A numerical fit to the 99.2~\% $^\mathrm{28}$Si distribution using a Lorentzian line shape leads to an indistinguishable agreement with the actual distribution (solid line).}
\end{figure}

In Fig.~\ref{figure2b}(a), the FWHM of the sum of the transition energy distributions of the outer $\Delta m$=1 transitions ($|1\rangle$$\leftrightarrow$$|2\rangle$ and $|3\rangle$$\leftrightarrow$$|4\rangle$) (circles), and the distribution of the $|2\rangle$$\leftrightarrow$$|3\rangle$ transition energies (diamonds) are shown as a function of the degree of isotopic purification. As an approximation, the ratio between the residual $^\mathrm{29}$Si and
$^\mathrm{30}$Si concentrations was kept constant at the
$^\mathrm{nat}$Si value. For both resonances, we observe a monotonous decrease of the isotope-induced broadening when the $^\mathrm{28}$Si concentration is increased. In Fig.~\ref{figure2b}(b), the line shape of the distributions of the outer $\Delta m$=1 transition energies are exemplarily shown for 99.2~\% (dashed line) and 99.9~\% (solid line) $^\mathrm{28}$Si. Both distributions are shown together with numerical fits assuming Gaussian lineshapes (dotted lines). For the 99.9~\% $^\mathrm{28}$Si distribution, a Lorentzian fit leads to a curve which cannot be distinguished from the actual distribution (solid line). Comparing the calculated distributions with the fits, it can be seen that the line shape of the distributions changes from Gaussian-like to Lorentzian in the regime of high isotopic purity with $>$99~\% $^\mathrm{28}$Si. This complicates the discrimination between isotope-induced and strain-induced broadening in the high $^\mathrm{28}$Si concentration range, where both types of inhomogeneous broadening are Lorentzian. For 99.98~\% $^\mathrm{28}$Si, the expected FWHM due to isotope effects for the $|2\rangle$$\leftrightarrow$$|3\rangle$ transition and for the sum of the $|1\rangle$$\leftrightarrow$$|2\rangle$ and
$|3\rangle$$\leftrightarrow$$|4\rangle$ distributions has decreased
to 0.002~$\mu$eV and 0.02~$\mu$eV, respectively. For the spectrum
shown in Fig.~\ref{figure3}(b), this means that the
isotope-related inhomogeneous contributions to the observed
linewidths are only 0.24~mT for the low-field resonance and
0.024~mT for the high-field line. These values fall more than one order of magnitude below the experimentally observed
linewidths. The linewidth of the broad $\Delta m$=1 resonance in
the spectra of our $^\mathrm{28}$Si sample should therefore only
be determined by spin relaxation and the concentration of point
defects inducing random strain fields. Under this assumption, we
can directly use Eq.~6 in Ref.~\onlinecite{Neubrand-pss90-301-1978} to
estimate the total concentration of C and O point defects from the
peak-to-peak linewidth at low temperatures. We find a value of
3$(\pm 2) \times$10$^{16}$~cm$^{-3}$, which is consistent with the $\approx$1$\times$10$^{16}$~cm$^{-3}$ concentration of C impurities measured by infrared absorption spectroscopy on the $\approx$607~cm$^{-1}$ C local vibrational mode.~\cite{Sennikov-SEMICOND-2005} A further reduction of the total point defect concentration to $\approx$1$\times$10$^{15}$~cm$^{-3}$
would be necessary to observe an isotope induced broadening in the
$^\mathrm{28}$Si sample investigated.

\begin{figure}
\includegraphics[width=7cm]{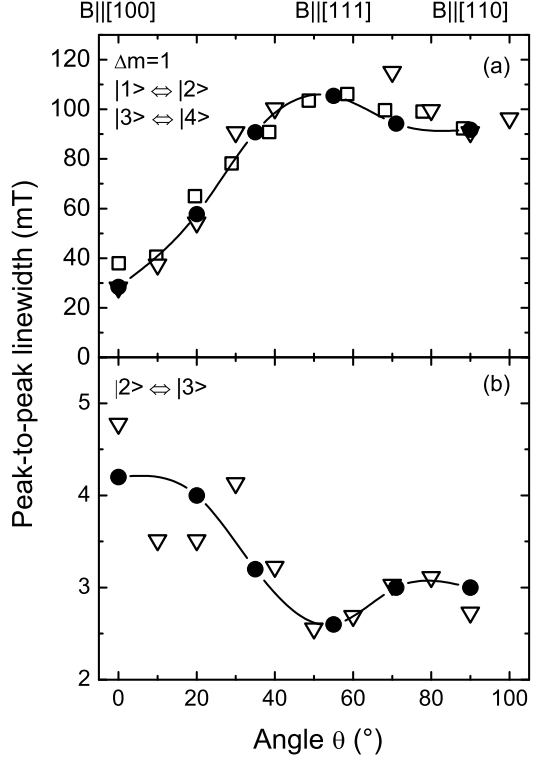}
\caption{\label{figure4} Linewidth anisotropy of the
$|1\rangle$$\leftrightarrow$$|2\rangle$,
$|3\rangle$$\leftrightarrow$$|4\rangle$ EPR transitions (a), and
the $|2\rangle$$\leftrightarrow$$|3\rangle$ transition, shown in
(b). The calculated angular dependence (full circles) for $^\mathrm{nat}$Si using $\Delta E^{29}$=0.74~meV and $\Delta E^{30}$=1.46~meV is compared
with experimentally determined linewidths for $^\mathrm{nat}$Si from this work (triangles)
and the anisotropy obtained from Fig.~4 in
Ref.~\onlinecite{Neubrand-pss86-269-1978} (squares).
The lines are guides to the eye.
}
\end{figure}

To obtain a more general picture, we have further investigated the
anisotropy of the isotope-induced broadening effects. We have calculated the broadening of the $\Delta m$=1 EPR lines
for different directions of the magnetic field ranging from
$B$$||$[001] ($\theta$=0$^\circ$) to $B$$||$[110]
($\theta$=90$^\circ$). In Fig.~\ref{figure4}(a), the peak-to-peak linewidth of the broad $\Delta m$=1 resonance
($|1\rangle$$\leftrightarrow$$|2\rangle$ and $|3\rangle$$\leftrightarrow$$|4\rangle$) calculated for $^\mathrm{nat}$Si using $\Delta E^{29}$=0.74~meV and $\Delta E^{30}$=1.46~meV is shown as a function of the angle $\theta$ (solid cirlces). We observe a strongly anisotropic behavior with a minimum of the
linewidth at $B$$||$[001] and a maximum at $B$$||$[111]. A
comparison with experimental data from
Ref.~\onlinecite{Neubrand-pss90-301-1978} (open squares) and those
obtained in this work (open triangles) shows a very good agreement
with our theoretical calculations. The experimental uncertainty
should be the smallest for small values of $\theta$ where narrow
lines with a large amplitude are observed. Regarding this, the experimental data from Ref.~\onlinecite{Neubrand-pss90-301-1978} show a comparatively
large deviation from our theoretical result for $\theta$=0$^\circ$
and $\theta$=20$^\circ$. These larger linewidths can
however be understood from a higher concentration of C and O point
defects in the sample measured by Neubrand, which imposes a lower limit to the experimentally observed linewidth via an additional inhomogeneous broadening induced by random local
strain.

In Fig.~\ref{figure4}(b), the theoretical and experimental data
obtained for the $|2\rangle$$\leftrightarrow$$|3\rangle$ resonance
are shown. Compared to Fig.~\ref{figure4}(a), the anisotropy is
inverted. We find a maximum of the linewidth for $B$$||$[001] and
a minimum for $B$$||$[111]. Again, the experimentally observed
linewidth anisotropy for $^\mathrm{nat}$Si is very nicely reproduced by our theoretical model.
The origin of the relatively large scatter of the experimental data points for $\theta$$<$30$^\circ$ is the
asymmetry of the $|2\rangle$$\leftrightarrow$$|3\rangle$ resonance
shown in Fig.~\ref{figure3}(a). For $B$$||$[001], the
positive amplitude $a$ (with respect to the zero crossing) on the
low-field side of the derivative line is smaller than the
negative amplitude $b$. This asymmetry of the line shape, which results from the mixing of the Zeeman states due to the isotopic perturbation, also
leads to a relatively weak and broad negative peak on the high field side,
which, for a given noise level, increases the experimental error
for the determination of the overall linewidth. In
Fig.~\ref{figure4}(b), we find the best agreement between theory
and experiment around $B$$||$[111], where we observe the smallest
linewidths.

\begin{figure}
\includegraphics[width=7cm]{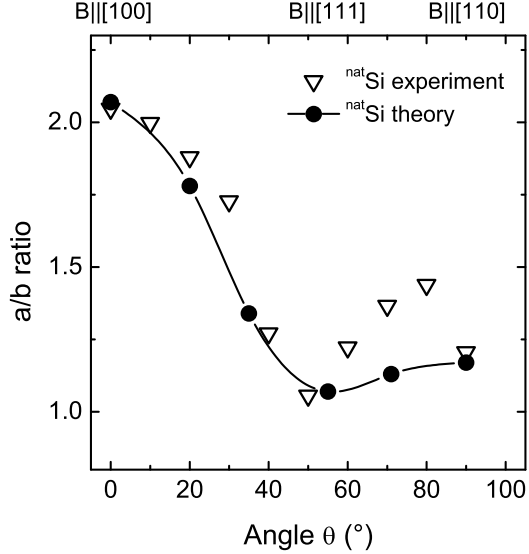}
\caption{\label{figure5}Angular dependence of the asymmetry ratio
$a/b$ of the $|2\rangle$$\leftrightarrow$$|3\rangle$ transition as
defined in Fig.~\ref{figure3}. The triangles show the
experimentally obtained values for the $^\mathrm{nat}$Si sample studied here, full circles represent the theoretical results.
The solid line is a guide to the eye.}
\end{figure}

As shown in Fig.~\ref{figure5}, also the asymmetry $a/b$ of the
$|2\rangle$$\leftrightarrow$$|3\rangle$ resonance line shape has a
minimum at this orientation.
Triangles show the experimental data, full circles represent the asymmetry that is deduced from the simulated
line shapes. With the exception of two data points around
$\theta$=70$^\circ$, the angular dependence of the line shape anisotropy can also quantitatively be understood from our theoretical
model. As mentioned above, the asymmetric broadening of the
$|2\rangle$$\leftrightarrow$$|3\rangle$ resonance originates from
a mixing of the unperturbed Zeeman states $|\Psi_i \rangle$ via
the off-diagonal matrix elements of the isotopic perturbation
Hamiltonian $\hat H_\mathrm{iso}^{4\times4}$. The difference between the experimental and the theoretical values of the asymmetry factor in the region of $\theta$=70$^\circ$ could be a result of a slight anisotropy of the linewidth of the single spin packets contributing to the inhomogeneously broadened line, which has not been taken into account for the simulation.
The scatter around $a/b$=1 in Fig.~\ref{figure5} lies within the experimental error bars resulting from the different spectral overlap of the resonance lines for different orientations of the magnetic field.
\begin{figure}
\includegraphics[width=8.5cm]{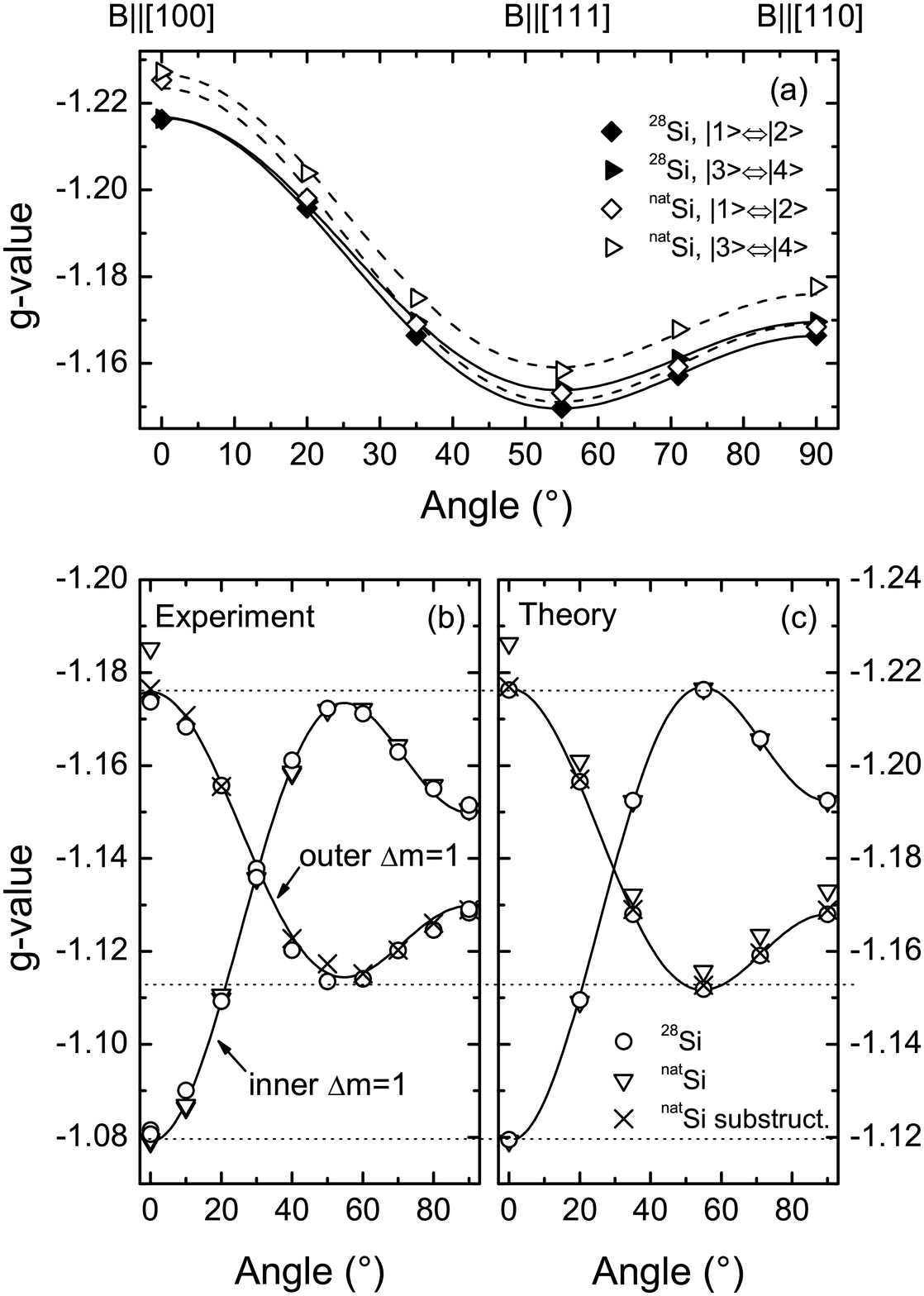}
\caption{\label{figure8n}(a) Theoretical anisotropy of the effective g-values of the outer $\Delta m$=1 transitions as obtained from our $\mathbf{k \cdot p}$ model for $^\mathrm{28}$Si (solid symbols) and including the isotopic perturbation in $^\mathrm{nat}$Si (open symbols). Diamonds and triangles indicate the g-values obtained for the $|1\rangle$$\leftrightarrow$$|2\rangle$ and $|3\rangle$$\leftrightarrow$$|4\rangle$ transitions, respectively. The solid and dashed lines are numerical fits to the data using the theoretical anisotropy given by Eq.~5 in Ref.~\onlinecite{Neubrand-pss86-269-1978} as a model.
In (b) and (c), a direct comparison of the experimental and the theoretical g-value anisotropies for both $\Delta m$=1 resonances is shown. Data points obtained for $^\mathrm{28}$Si are indicated by open circles,  $^\mathrm{nat}$Si data points are shown as open triangles. The theoretical data for the outer $\Delta m$=1 resonance is a weighted average of the $|1\rangle$$\leftrightarrow$$|2\rangle$ and $|3\rangle$$\leftrightarrow$$|4\rangle$ transitions shown in (a) for $T$=3~K. The crosses show the g-values of the substructure line. Details concerning the calculations of the data shown in (c) are described in the text.}
\end{figure}

The calculated distributions of energy splittings in Fig.~\ref{figure2} and the experimental EPR spectra in Figs.~\ref{figure3} and Fig.~\ref{figure3b} revealed that the maxima of the distributions of transition energies in $^\mathrm{nat}$Si are shifted with respect to the corresponding transition energies observed for $^\mathrm{28}$Si. In Fig.~\ref{figure8n}(a), we show how this effect depends on the orientation of the magnetic field for the outer $\Delta m$=1 transitions.
The effective g-values calculated from the theoretical $|1\rangle$$\leftrightarrow$$|2\rangle$ and $|3\rangle$$\leftrightarrow$$|4\rangle$ transition energies in pure $^\mathrm{28}$Si are shown as full diamonds and triangles, respectively. The open symbols indicate the corresponding effective g-values for $^\mathrm{nat}$Si calculated from the maxima of the transition energy distributions. According to Ref.~\onlinecite{Bir-JPCS-1963}, we assume the g-values of B in Si to be negative. As guides to the eye, we also show numerical fits to the data based on the theoretical anisotropy given by Eq.~5 in Ref.~\onlinecite{Neubrand-pss86-269-1978} as solid and dashed lines. For $^\mathrm{28}$Si, we observe a splitting in energy of the two outer $\Delta m$=1 transitions for $\theta > 0^\circ$. This splitting has a maximum for $B$$||$[111] and should result in a splitting of the broad $\Delta m$=1 EPR resonance. However, we cannot resolve this splitting in our experiments due to the comparatively strong defect-induced inhomogeneous broadening. For $^\mathrm{nat}$Si, the g-values are shifted to more negative values with respect to $^\mathrm{28}$Si for both transitions. The $|3\rangle$$\leftrightarrow$$|4\rangle$ resonance is shifted more strongly than the $|1\rangle$$\leftrightarrow$$|2\rangle$ resonance.

In our EPR experiments we observe only one resonance with a g-value that corresponds to an average of the g-values of the two outer $\Delta m$=1 transitions that is weighted according to the equilibrium populations of the different spin states.
In Fig.~\ref{figure8n}(b), the experimentally obtained g-value anisotropies of the outer- as well as the inner $\Delta m$=1 resonances for $^\mathrm{28}$Si (open circles) and $^\mathrm{nat}$Si (open triangles) are plotted. In addition, also the effective g-values of the substructure superimposing the outer $\Delta m$=1 resonance are shown for $^\mathrm{nat}$Si (crosses).
For direct comparison, the corresponding theoretical results are shown in Fig.~\ref{figure8n}(c). The effective g-values of the outer $\Delta m$=1 resonances were calculated by taking the maxima of the weighted sums of the $|1\rangle$$\leftrightarrow$$|2\rangle$ and $|3\rangle$$\leftrightarrow$$|4\rangle$ transition energy distributions assuming Boltzmann distributed populations of the four spin states ($T$=3~K).
The effective g-values at which the substructure line on the broad $\Delta m$=1 resonance is expected were calculated under the assumption that the substructure lines result from a subensemble of acceptors for which the transition energies of the two outer $\Delta m$=1 transitions are equal, i.e.~the energy level scheme of the four level spin system is symmetric.~\cite{Neubrand-pss86-269-1978} The solid lines in Figs.~\ref{figure8n}(b) and (c) are again numerical fits using the model described in Ref.~\onlinecite{Neubrand-pss86-269-1978}.
We note that the g-value scale in Fig.~\ref{figure8n}(c) is shifted by 0.04 with respect to the one in Fig.~\ref{figure8n}(b) to enable a better comparison between the experimental and theoretical data. Disregarding this rigid shift, which is caused by the error of the absolute g-values obtained from our $\mathbf{k \cdot p}$ model, our calculations can be used to describe and understand the different experimental observations.
For the inner $\Delta m$=1 transition $|2\rangle$$\leftrightarrow$$|3\rangle$, we could already see in Figs.~\ref{figure2} and \ref{figure3} that the isotopic perturbation does not lead to a measurable shift in the g-value for the 0$^\circ$ orientation. As can be seen in Fig.~\ref{figure8n}, this finding holds for all orientations of the magnetic field. Comparing the anisotropies of the outer $\Delta m$=1 resonance in Figs.~\ref{figure8n}(b) and (c) for $^\mathrm{28}$Si and $^\mathrm{nat}$Si, we observe a similar shift of the g-values. Due to the large linewidth of the latter resonance of up to 100~mT in $^\mathrm{nat}$Si, it was not possible to extract a reliable g-value for all intermediate orientations of the magnetic field from our experimental spectra.

The g-values of the substructure line in $^\mathrm{nat}$Si, which could be determined for all orientations of the magnetic field and with a higher precision due to its comparatively small linewidth, show only a small deviation from the effective g-values of the outer $\Delta m$=1 resonance in $^\mathrm{28}$Si irrespective of $\theta$. In clear contrast to what has been reported in Ref.~\onlinecite{Neubrand-pss86-269-1978}, this means that the resonance field of the substructure line does not coincide with the center of the broad $\Delta m$=1 resonance in $^\mathrm{nat}$Si. The agreement between our theoretical model and the experimental data in this point strongly indicates that the substructure lines indeed originate from a subensemble of acceptors where the energy splittings between the states $|1\rangle$ and $|2\rangle$ and the states $|3\rangle$ and $|4\rangle$ are equal.

For the $\Delta m$=1 substructure line in $^\mathrm{nat}$Si, the best fits to our experimental data is obtained for $g_1$=-1.0776 and $g_2$=-0.0307. For the inner $\Delta m$=1 resonance, we obtain $g_1$=-1.0728 and $g_2$=-0.0315. In Ref.~\onlinecite{Neubrand-pss86-269-1978}, for comparison, $g_1$=-1.0740 and $g_2$=-0.0307 was determined for the $\Delta m$=1 substructure and $g_1$=-1.0676 and $g_2$=-0.0317 was measured for the inner $\Delta m$=1 resonance. We find a good agreement of our data with the original reports in particular for $g_2$, which determines the angular dependence of the effective g-values. For $g_1$, which causes an angle-independent offset of the g-value anisotropy, we obtain slightly lower values. This deviation might result from differences in the calibration of the magnetic field.

\begin{figure}
\includegraphics[width=7cm]{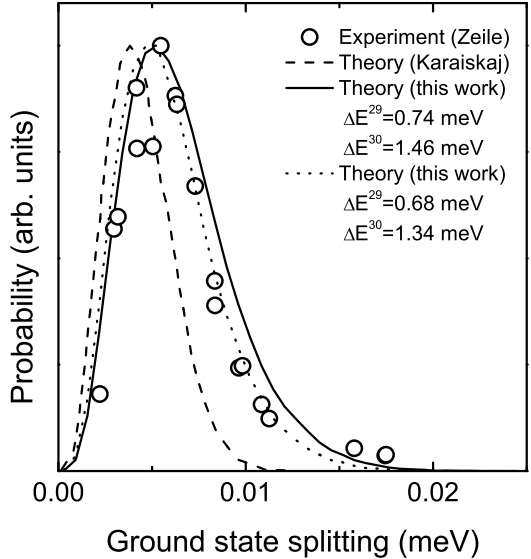}
\caption{\label{figure6}Energy distribution of the residual B
acceptor ground state splitting in the absence of an external
magnetic field. The open circles show the combination of two sets of experimental data measured with Si samples with a boron concentration of 5.4$\times$10$^{15}$~cm$^{-3}$ and 2$\times$10$^{14}$~cm$^{-3}$. The data is taken
from Fig.~3 in Ref.~\onlinecite{Zeile-pss-1982}. The dashed line
represents the distribution calculated in
Ref.~\onlinecite{Karaiskaj-PRL90-016404-2003}. The solid line
shows the results obtained with our theoretical method using the
same values for the valence band offsets between the different Si
isotopes as in Ref.~\onlinecite{Karaiskaj-PRL90-016404-2003}, and
the dotted line is obtained when $\Delta E^\mathrm{29}$ and $\Delta E^\mathrm{30}$ are
decreased by 8~\%}
\end{figure}

Figures~\ref{figure3} to \ref{figure5} show that the line shape and width of the $\Delta m$=1 and $\Delta m$=2 resonances in $^\mathrm{nat}$Si with a low concentration of point defects can quantitatively be understood by our model accounting for the fluctuations of the valence band position caused by the disorder in the placement of the different Si isotopes in the crystal lattice in the vicinity of the different B acceptor nuclei.
In contrast, the original model used to calculate the acceptor
ground state splitting for $B$=0 could not explain the
experimentally observed value
quantitatively.~\cite{Karaiskaj-PRL90-016404-2003} In
Fig.~\ref{figure6}, the calculated statistical distribution of the
B acceptor ground state splitting taken from
Ref.~\onlinecite{Karaiskaj-PRL90-016404-2003} (dashed line) is
shown together with experimental data that were obtained from
phonon absorption spectroscopy.~\cite{Zeile-pss-1982,Zeile-PSS-2-1982}. The shapes of the distributions are in good agreement, however, the maximum of the
splitting energy shows a discrepancy of 30~\%. To investigate the
origin of this discrepancy, we have repeated the calculations for the distribution of the residual ground state splitting in $^\mathrm{nat}$Si
applying our theoretical method to $B$=0. The result is depicted
as the solid line in Fig.~\ref{figure6} and shows a much better
quantitative agreement with the experimental data. Besides the
potential $Q$ at the central grid node, which was not adjusted for
the calculations for $B$=0, the most relevant parameters in our
model are the valence band offsets between the different Si
isotopes ($V_\mathrm{iso}$). Since the values used for these
parameters in this work so far are identical to the ones used in
Ref.~\onlinecite{Karaiskaj-PRL90-016404-2003}, the discrepancy
between the acceptor ground state splitting at $B$=0 calculated in
the original work and in the present work can not be
attributed to an uncertainty in the valence band offsets deduced
from Ref.~\onlinecite{Cardona-PEP-1989}. It should rather result
from a less realistic modeling of the boron acceptor wave
function. In order to see how sensitive the residual ground state
splitting depends on the assumed valence band offsets, we have
recalculated the distribution of splitting energies for different
$\Delta E^\mathrm{29}$. The second parameter $\Delta
E^\mathrm{30}$ is directly linked to the value of $\Delta E^\mathrm{29}$ by the $M^{-1/2}$ dependence of the
renormalization energy on the isotope mass $M$. Using our wave functions, the best overall agreement with the experimental data shown in
Fig.~\ref{figure6} is obtained when $\Delta E^\mathrm{29}$ and $\Delta E^\mathrm{30}$ are decreased by
8~\% from their original value (dotted line).
We have also recalculated the EPR spectrum of the $\Delta m$=1 resonances in $^\mathrm{nat}$Si with the reduced $\Delta E$ values and the result of this simulation is shown as the dashed line in Fig.~\ref{figure3}(a). Using the lower band offsets, we obtain a slightly worse agreement with the experimental spectrum on the low-field edge of the broad $|1\rangle$$\leftrightarrow$$|2\rangle$, $|3\rangle$$\leftrightarrow$$|4\rangle$ resonance, while an equally good or even slightly better agreement is found for the central region and the high field edge of this resonance as well as for the $|2\rangle$$\leftrightarrow$$|3\rangle$ line.
In view of the overall agreement with the experimental data and neglecting possible inaccuracies of the wave functions obtained from our calculations, we suggest a slight correction of the valence band offsets between the different Si isotopes to $\Delta E^\mathrm{29}$=0.68~meV and $\Delta E^\mathrm{29}$=1.34~meV.
We note that a deviation from the original
values by more than 5~\% to higher energies and more than 10~\% to
lower energies would lead to a significant discrepancy between
theory and experiment.

\section{Conclusions}
We have set up a theoretical model that allows for a quantitative
description of the isotopic effect on the Zeeman energy level
scheme of paramagnetic B acceptors in Si, accounting for the
isotopic perturbation via local fluctuations of the valence band
edge at $^\mathrm{29}$Si and $^\mathrm{30}$Si lattice sites. Using
this model, we can reproduce the experimentally observed line
shapes of the B-related $\Delta m$=1 and $\Delta m$=2 resonances in $^\mathrm{nat}$Si with excellent agreement between theory and experiment. Our results demonstrate
that the inhomogeneous broadening of B-related EPR lines measured
in ultra-pure natural silicon
can quantitatively be understood from the isotopic perturbation,
both in its magnitude as well as in the resulting line shapes, without the necessity to invoke hyperfine interaction with the B nucleus or with $^\mathrm{29}$Si.
Concretely, the open questions raised in the introduction can be
answered as follows: (i) As can directly be deduced from a
comparison of the data shown in Fig.~\ref{figure4} and Fig.~2 in
Ref.~\onlinecite{Neubrand-pss90-301-1978}, the Gaussian
contribution to the line shape of the broad $\Delta m$=1 resonance
is merely a phenomenological way to account for the broadening
induced by the isotopic randomness. However, a Gaussian line is no
realistic representation of the isotope-induced broadening, which
is determined by the distribution of transition energies shown in
Fig.~\ref{figure2}. Therefore, an additional Lorentz line was
needed in Ref.~\onlinecite{Neubrand-pss90-301-1978} to obtain a
reasonable fit of the broad $\Delta m$=1 resonance measured with
the samples of highest crystalline purity. Consequently, the 10~mT
threshold of the Lorentzian contribution to the line shape (ii) was an
artifact that resulted from an inappropriate model of the line
shape in the limit of low point defect concentrations.
We have found that the perturbation due to the random distribution of the different Si isotopes also leads to a shift in the effective g-values, mainly of the broad $\Delta m$=1 and the $\Delta m$=2 resonances. We note that this isotope induced shift could in principle be used to control the g-values of a B acceptor by tailoring its isotopic surrounding. This could also be achieved dynamically by placing an acceptor close to an isotopic heterojunction and manipulating its wave function with the help of electric fields. This idea might particularly be of interest for applications e.g.~in quantum computation technology, as it could be realized in a completely nuclear spin free environment only using $^\mathrm{28}$Si and $^\mathrm{30}$Si isotopes. To give an example, we obtain an average g-value of the broad $\Delta m$=1 resonance of -1.267 for an isotopically engineered material composed of 50~\% $^\mathrm{28}$Si and 50~\% $^\mathrm{30}$Si ($B || [100]$, $T$=3~K), which corresponds to a 4.2~\% change of the g-value compared to $g$=-1.216 calculated for pure $^\mathrm{28}$Si.
Concerning issue (iii), our results suggest that the substructure lines of the $\Delta m$=1 and $\Delta m$=2 resonances originate from a subensemble of acceptors, where the Zeeman level scheme is perturbed such that the $|1\rangle$$\leftrightarrow$$|2\rangle$, $|3\rangle$$\leftrightarrow$$|4\rangle$ transition energies are equal. We discuss these substructure lines and their dynamics in more detail in Ref.~\onlinecite{Tezuka-2009}. Finally, a comparison of our calculation with previous works
investigating the B acceptor ground state in the absence of an
external magnetic field, provides an independent verification of
the energy offsets between the valence bands of the different
isotopes of silicon.

\begin{acknowledgments}
Work at Walter Schottky Institut has been supported by the DFG
(Grants No. SFB 631, C1 and C3, and Br 1585/5). Further, we acknowledge support from the JST-DFG Strategic Cooperative Program on Nanoelectronics, the Grant-in-Aid
for Scientific Research \#18001002,  Special Coordination Funds
for Promoting Science and Technology, and a Grant-in-Aid for the
Global Center of Excellence at Keio University.
\end{acknowledgments}


\begin{thebibliography}{25}
\expandafter\ifx\csname natexlab\endcsname\relax\def\natexlab#1{#1}\fi
\expandafter\ifx\csname bibnamefont\endcsname\relax
  \def\bibnamefont#1{#1}\fi
\expandafter\ifx\csname bibfnamefont\endcsname\relax
  \def\bibfnamefont#1{#1}\fi
\expandafter\ifx\csname citenamefont\endcsname\relax
  \def\citenamefont#1{#1}\fi
\expandafter\ifx\csname url\endcsname\relax
  \def\url#1{\texttt{#1}}\fi
\expandafter\ifx\csname urlprefix\endcsname\relax\def\urlprefix{URL }\fi
\providecommand{\bibinfo}[2]{#2}
\providecommand{\eprint}[2][]{\url{#2}}

\bibitem[{\citenamefont{Feher}(1959)}]{Feher-PRB-I-1959}
\bibinfo{author}{\bibfnamefont{G.}~\bibnamefont{Feher}},
  \bibinfo{journal}{Phys. Rev.} \textbf{\bibinfo{volume}{114}},
  \bibinfo{pages}{1219} (\bibinfo{year}{1959}).

\bibitem[{\citenamefont{Feher and Gere}(1959)}]{Feher-PRB-II-1959}
\bibinfo{author}{\bibfnamefont{G.}~\bibnamefont{Feher}} \bibnamefont{and}
  \bibinfo{author}{\bibfnamefont{E.~A.} \bibnamefont{Gere}},
  \bibinfo{journal}{Phys. Rev.} \textbf{\bibinfo{volume}{114}},
  \bibinfo{pages}{1245} (\bibinfo{year}{1959}).

\bibitem[{\citenamefont{Wilson and Feher}(1961)}]{Wilson-PRB-1961}
\bibinfo{author}{\bibfnamefont{D.~K.} \bibnamefont{Wilson}} \bibnamefont{and}
  \bibinfo{author}{\bibfnamefont{G.}~\bibnamefont{Feher}},
  \bibinfo{journal}{Phys. Rev.} \textbf{\bibinfo{volume}{124}},
  \bibinfo{pages}{1068} (\bibinfo{year}{1961}).

\bibitem[{\citenamefont{Stutzmann et~al.}(1987)\citenamefont{Stutzmann,
  Biegelsen, and Street}}]{Stutzmann-PRB-1987}
\bibinfo{author}{\bibfnamefont{M.}~\bibnamefont{Stutzmann}},
  \bibinfo{author}{\bibfnamefont{D.~K.} \bibnamefont{Biegelsen}},
  \bibnamefont{and} \bibinfo{author}{\bibfnamefont{R.~A.}
  \bibnamefont{Street}}, \bibinfo{journal}{Phys. Rev. B}
  \textbf{\bibinfo{volume}{35}}, \bibinfo{pages}{5666} (\bibinfo{year}{1987}).

\bibitem[{\citenamefont{M\"{u}ller et~al.}(1999)\citenamefont{M\"{u}ller,
  Finger, Carius, and Wagner}}]{Mueller-PRB-1999}
\bibinfo{author}{\bibfnamefont{J.}~\bibnamefont{M\"{u}ller}},
  \bibinfo{author}{\bibfnamefont{F.}~\bibnamefont{Finger}},
  \bibinfo{author}{\bibfnamefont{R.}~\bibnamefont{Carius}}, \bibnamefont{and}
  \bibinfo{author}{\bibfnamefont{H.}~\bibnamefont{Wagner}},
  \bibinfo{journal}{Phys. Rev. B} \textbf{\bibinfo{volume}{60}},
  \bibinfo{pages}{11666} (\bibinfo{year}{1999}).

\bibitem[{\citenamefont{Huebl et~al.}(2006)\citenamefont{Huebl, Stegner,
  Stutzmann, Brandt, Vogg, Bensch, and Gerstmann}}]{Huebl-PRL-2006}
\bibinfo{author}{\bibfnamefont{H.}~\bibnamefont{Huebl}},
  \bibinfo{author}{\bibfnamefont{A.~R.} \bibnamefont{Stegner}},
  \bibinfo{author}{\bibfnamefont{M.}~\bibnamefont{Stutzmann}},
  \bibinfo{author}{\bibfnamefont{M.~S.} \bibnamefont{Brandt}},
  \bibinfo{author}{\bibfnamefont{G.}~\bibnamefont{Vogg}},
  \bibinfo{author}{\bibfnamefont{F.}~\bibnamefont{Bensch}}, \bibnamefont{and}
  \bibinfo{author}{\bibfnamefont{U.}~\bibnamefont{Gerstmann}},
  \bibinfo{journal}{Phys. Rev. Lett.} \textbf{\bibinfo{volume}{97}},
  \bibinfo{pages}{166402} (\bibinfo{year}{2006}).

\bibitem[{\citenamefont{Pereira et~al.}(2009)\citenamefont{Pereira, Stegner,
  Andlauer, Klein, Wiggers, Brandt, and Stutzmann}}]{Pereira-PRB-2009}
\bibinfo{author}{\bibfnamefont{R.~N.} \bibnamefont{Pereira}},
  \bibinfo{author}{\bibfnamefont{A.~R.} \bibnamefont{Stegner}},
  \bibinfo{author}{\bibfnamefont{T.}~\bibnamefont{Andlauer}},
  \bibinfo{author}{\bibfnamefont{K.}~\bibnamefont{Klein}},
  \bibinfo{author}{\bibfnamefont{H.}~\bibnamefont{Wiggers}},
  \bibinfo{author}{\bibfnamefont{M.~S.} \bibnamefont{Brandt}},
  \bibnamefont{and}
  \bibinfo{author}{\bibfnamefont{M.}~\bibnamefont{Stutzmann}},
  \bibinfo{journal}{Phys. Rev. B} \textbf{\bibinfo{volume}{79}},
  \bibinfo{pages}{161304(R)} (\bibinfo{year}{2009}).

\bibitem[{\citenamefont{Kohn}(1957)}]{Kohn-SSP-1957}
\bibinfo{author}{\bibfnamefont{W.}~\bibnamefont{Kohn}},
  \emph{\bibinfo{title}{Solid State Physics Vol. 5}}, \bibnamefont{edited by F. Seitz and D. Turnbull}
  (\bibinfo{publisher}{Academic Press, New York}, \bibinfo{year}{1957}), p.
  \bibinfo{pages}{257}.

\bibitem[{\citenamefont{Feher et~al.}(1960)\citenamefont{Feher, Hensel, and
  Gere}}]{Feher-PRL-I-1960}
\bibinfo{author}{\bibfnamefont{G.}~\bibnamefont{Feher}},
  \bibinfo{author}{\bibfnamefont{J.~C.} \bibnamefont{Hensel}},
  \bibnamefont{and} \bibinfo{author}{\bibfnamefont{E.~A.} \bibnamefont{Gere}},
  \bibinfo{journal}{Phys. Rev. Lett.} \textbf{\bibinfo{volume}{5}},
  \bibinfo{pages}{309} (\bibinfo{year}{1960}).

\bibitem[{\citenamefont{Neubrand}(1978{\natexlab{a}})}]{Neubrand-pss86-269-197%
8}
\bibinfo{author}{\bibfnamefont{H.}~\bibnamefont{Neubrand}},
  \bibinfo{journal}{phys. stat. sol. (b)} \textbf{\bibinfo{volume}{86}},
  \bibinfo{pages}{269} (\bibinfo{year}{1978}{\natexlab{a}}).

\bibitem[{\citenamefont{Neubrand}(1978{\natexlab{b}})}]{Neubrand-pss90-301-197%
8}
\bibinfo{author}{\bibfnamefont{H.}~\bibnamefont{Neubrand}},
  \bibinfo{journal}{phys. stat. sol. (b)} \textbf{\bibinfo{volume}{90}},
  \bibinfo{pages}{301} (\bibinfo{year}{1978}{\natexlab{b}}).

\bibitem[{\citenamefont{K\"{o}pf and Lassmann}(1992)}]{Lassmann-PRL-1992}
\bibinfo{author}{\bibfnamefont{A.}~\bibnamefont{K\"{o}pf}} \bibnamefont{and}
  \bibinfo{author}{\bibfnamefont{K.}~\bibnamefont{Lassmann}},
  \bibinfo{journal}{Phys. Rev. Lett.} \textbf{\bibinfo{volume}{69}},
  \bibinfo{pages}{1580} (\bibinfo{year}{1992}).

\bibitem[{\citenamefont{Karaiskaj et~al.}(2002)\citenamefont{Karaiskaj,
  Thewalt, Ruf, Cardona, and Konuma}}]{Karaiskaj-PRL89-016401-2002}
\bibinfo{author}{\bibfnamefont{D.}~\bibnamefont{Karaiskaj}},
  \bibinfo{author}{\bibfnamefont{M.~L.~W.} \bibnamefont{Thewalt}},
  \bibinfo{author}{\bibfnamefont{T.}~\bibnamefont{Ruf}},
  \bibinfo{author}{\bibfnamefont{M.}~\bibnamefont{Cardona}}, \bibnamefont{and}
  \bibinfo{author}{\bibfnamefont{M.}~\bibnamefont{Konuma}},
  \bibinfo{journal}{Phys. Rev. Lett.} \textbf{\bibinfo{volume}{89}},
  \bibinfo{pages}{016401} (\bibinfo{year}{2002}).

\bibitem[{\citenamefont{Karaiskaj et~al.}(2003)\citenamefont{Karaiskaj,
  Kirczenow, Thewalt, Buczko, and Cardona}}]{Karaiskaj-PRL90-016404-2003}
\bibinfo{author}{\bibfnamefont{D.}~\bibnamefont{Karaiskaj}},
  \bibinfo{author}{\bibfnamefont{G.}~\bibnamefont{Kirczenow}},
  \bibinfo{author}{\bibfnamefont{M.~L.~W.} \bibnamefont{Thewalt}},
  \bibinfo{author}{\bibfnamefont{R.}~\bibnamefont{Buczko}}, \bibnamefont{and}
  \bibinfo{author}{\bibfnamefont{M.}~\bibnamefont{Cardona}},
  \bibinfo{journal}{Phys. Rev. Lett.} \textbf{\bibinfo{volume}{90}},
  \bibinfo{pages}{016404} (\bibinfo{year}{2003}).

\bibitem[{\citenamefont{Cardona and Gopalan}(1989)}]{Cardona-PEP-1989}
\bibinfo{author}{\bibfnamefont{M.}~\bibnamefont{Cardona}} \bibnamefont{and}
  \bibinfo{author}{\bibfnamefont{S.}~\bibnamefont{Gopalan}},
  \emph{\bibinfo{title}{Progress in Electron Properties of Solids}},  \bibnamefont{edited by R. Girlanda} \emph{et al.}
  (\bibinfo{publisher}{Kluwer, Dordrecht}, \bibinfo{year}{1989}),
  p.~\bibinfo{pages}{51}.

\bibitem[{\citenamefont{Birner et~al.}(2007)\citenamefont{Birner, Zibold,
  Andlauer, Kubis, Sabathil, Trellakis, and Vogl}}]{Birner-IEEE-2007}
\bibinfo{author}{\bibfnamefont{S.}~\bibnamefont{Birner}},
  \bibinfo{author}{\bibfnamefont{T.}~\bibnamefont{Zibold}},
  \bibinfo{author}{\bibfnamefont{T.}~\bibnamefont{Andlauer}},
  \bibinfo{author}{\bibfnamefont{T.}~\bibnamefont{Kubis}},
  \bibinfo{author}{\bibfnamefont{M.}~\bibnamefont{Sabathil}},
  \bibinfo{author}{\bibfnamefont{A.}~\bibnamefont{Trellakis}},
  \bibnamefont{and} \bibinfo{author}{\bibfnamefont{P.}~\bibnamefont{Vogl}},
  \bibinfo{journal}{IEEE Trans. Electron Devices}
  \textbf{\bibinfo{volume}{54}}, \bibinfo{pages}{2137} (\bibinfo{year}{2007}).

\bibitem[{\citenamefont{Pantelides and Sah}(1974)}]{Pantelides-PRB-1974}
\bibinfo{author}{\bibfnamefont{S.~T.} \bibnamefont{Pantelides}}
  \bibnamefont{and} \bibinfo{author}{\bibfnamefont{C.~T.} \bibnamefont{Sah}},
  \bibinfo{journal}{Phys. Rev. B} \textbf{\bibinfo{volume}{10}},
  \bibinfo{pages}{621} (\bibinfo{year}{1974}).

\bibitem[{\citenamefont{Belyakov and Burdov}(2007)}]{Belyakov-JPHYS-2007}
\bibinfo{author}{\bibfnamefont{V.~A.} \bibnamefont{Belyakov}} \bibnamefont{and}
  \bibinfo{author}{\bibfnamefont{V.~A.} \bibnamefont{Burdov}},
  \bibinfo{journal}{J. Phys.: Condens. Matter} \textbf{\bibinfo{volume}{20}},
  \bibinfo{pages}{025213} (\bibinfo{year}{2007}).

\bibitem[{\citenamefont{Tezuka et~al.}(2010)\citenamefont{Tezuka, Stegner,
  Tyryshkin, Shankar, Thewalt, Lyon, Itoh, and Brandt}}]{Tezuka-2009}
\bibinfo{author}{\bibfnamefont{H.}~\bibnamefont{Tezuka}},
  \bibinfo{author}{\bibfnamefont{A.~R.} \bibnamefont{Stegner}},
  \bibinfo{author}{\bibfnamefont{A.~M.} \bibnamefont{Tyryshkin}},
  \bibinfo{author}{\bibfnamefont{S.}~\bibnamefont{Shankar}},
  \bibinfo{author}{\bibfnamefont{M.~L.~W.} \bibnamefont{Thewalt}},
  \bibinfo{author}{\bibfnamefont{S.~A.} \bibnamefont{Lyon}},
  \bibinfo{author}{\bibfnamefont{K.~M.} \bibnamefont{Itoh}}, \bibnamefont{and}
  \bibinfo{author}{\bibfnamefont{M.~S.} \bibnamefont{Brandt}},
  \bibinfo{journal}{arXiv:1003.4339v1}  (\bibinfo{year}{2010}).

\bibitem[{\citenamefont{Bir et~al.}(1963)\citenamefont{Bir, Butikov, and
  Pikus}}]{Bir-JPCS-1963}
\bibinfo{author}{\bibfnamefont{G.~L.} \bibnamefont{Bir}},
  \bibinfo{author}{\bibfnamefont{E.~I.} \bibnamefont{Butikov}},
  \bibnamefont{and} \bibinfo{author}{\bibfnamefont{G.~E.} \bibnamefont{Pikus}},
  \bibinfo{journal}{J. Phys. Chem. Solids} \textbf{\bibinfo{volume}{24}},
  \bibinfo{pages}{1467} (\bibinfo{year}{1963}).

\bibitem[{\citenamefont{Cardona and Thewalt}(2005)}]{Cardona-RMP-2005}
\bibinfo{author}{\bibfnamefont{M.}~\bibnamefont{Cardona}} \bibnamefont{and}
  \bibinfo{author}{\bibfnamefont{M.~L.~W.} \bibnamefont{Thewalt}},
  \bibinfo{journal}{Rev. Mod. Phys.} \textbf{\bibinfo{volume}{77}},
  \bibinfo{pages}{1173} (\bibinfo{year}{2005}).

\bibitem[{\citenamefont{Fischer et~al.}(2008)\citenamefont{Fischer, Coish,
  Bulaev, and Loss}}]{Fischer-ARXIV-2008}
\bibinfo{author}{\bibfnamefont{J.}~\bibnamefont{Fischer}},
  \bibinfo{author}{\bibfnamefont{W.~A.} \bibnamefont{Coish}},
  \bibinfo{author}{\bibfnamefont{D.~V.} \bibnamefont{Bulaev}},
  \bibnamefont{and} \bibinfo{author}{\bibfnamefont{D.}~\bibnamefont{Loss}},
  \bibinfo{journal}{Phys. Rev. B} \textbf{\bibinfo{volume}{78}},
  \bibinfo{pages}{155329} (\bibinfo{year}{2008}).

\bibitem[{\citenamefont{Sennikov et~al.}(2005)\citenamefont{Sennikov, Kotereva,
  Kurganov, Andreev, Niemann, Schiel, Emtsev, and
  Pohl}}]{Sennikov-SEMICOND-2005}
\bibinfo{author}{\bibfnamefont{P.}~\bibnamefont{Sennikov}},
  \bibinfo{author}{\bibfnamefont{T.}~\bibnamefont{Kotereva}},
  \bibinfo{author}{\bibfnamefont{A.}~\bibnamefont{Kurganov}},
  \bibinfo{author}{\bibfnamefont{B.}~\bibnamefont{Andreev}},
  \bibinfo{author}{\bibfnamefont{H.}~\bibnamefont{Niemann}},
  \bibinfo{author}{\bibfnamefont{D.}~\bibnamefont{Schiel}},
  \bibinfo{author}{\bibfnamefont{V.}~\bibnamefont{Emtsev}}, \bibnamefont{and}
  \bibinfo{author}{\bibfnamefont{H.~J.} \bibnamefont{Pohl}},
  \bibinfo{journal}{Semiconductors} \textbf{\bibinfo{volume}{39}},
  \bibinfo{pages}{300} (\bibinfo{year}{2005}).

\bibitem[{\citenamefont{Zeile and Lassmann}(1982)}]{Zeile-pss-1982}
\bibinfo{author}{\bibfnamefont{H.}~\bibnamefont{Zeile}} \bibnamefont{and}
  \bibinfo{author}{\bibfnamefont{K.}~\bibnamefont{Lassmann}},
  \bibinfo{journal}{phys. stat. sol. (b)} \textbf{\bibinfo{volume}{111}},
  \bibinfo{pages}{555} (\bibinfo{year}{1982}).

\bibitem[{\citenamefont{Zeile et~al.}(1982)\citenamefont{Zeile, Harten, and
  Lassmann}}]{Zeile-PSS-2-1982}
\bibinfo{author}{\bibfnamefont{H.}~\bibnamefont{Zeile}},
  \bibinfo{author}{\bibfnamefont{U.}~\bibnamefont{Harten}}, \bibnamefont{and}
  \bibinfo{author}{\bibfnamefont{K.}~\bibnamefont{Lassmann}},
  \bibinfo{journal}{phys. stat. sol. (b)} \textbf{\bibinfo{volume}{111}},
  \bibinfo{pages}{213} (\bibinfo{year}{1982}).

\end{thebibliography}
\end{document}